\documentclass[aps,reprint]{revtex4-1}
\usepackage{amsmath}
\usepackage{amssymb}
\usepackage{graphicx}
\usepackage{placeins}
\usepackage{caption}
\usepackage{subcaption}
\usepackage{hyperref}
\usepackage{braket}
\usepackage{cleveref}
\usepackage{mathrsfs}
\usepackage{graphicx}
\usepackage{dcolumn}
\usepackage{bm}
\usepackage{hyperref}
\usepackage[mathlines]{lineno}
\usepackage{color}
\begin{abstract}
We develop a fast, accurate, and robust technique to model the electromagnetic response of a two-dimensional (2D) waveguide-fed metasurface aperture. The geometry under consideration consists of a parallel-plate waveguide with an array of subwavelength, complementary metamaterial elements patterned into one of the conducting surfaces. To determine the radiated field, we model each radiating element as a polarizable dipole, and account for mutual interactions among elements via guided and radiated fields by applying a coupled dipole framework. Using full-wave simulations of parallel-plate waveguides with two types of metamaterial elements, we confirm the validity of the proposed coupled dipole model and demonstrate the ability to predict radiation patterns corresponding to arbitrarily arranged elements. We explore the importance of including the mutual coupling among metamaterial elements to arrive at accurate field predictions. The coupled dipole modeling framework presented is scalable to extremely large apertures and can form the foundation for a general, efficient, yet simple aperture analysis and synthesis tool.
\end{abstract}

\begin{document}
\title{Analytical Modeling of a Two-Dimensional Waveguide-Fed Metasurface}
\author{Laura Pulido Mancera$^{1}$, Mohammadreza F. Imani $^{1}$, Patrick T. Bowen $^{1}$, Nathan Kundtz $^{2}$, David R. Smith $^{1}$}
\affiliation{$^{1}$Department of Electrical and Computer Engineering, Duke University, Durham, NC 27708\\ $^{2}$Kymeta Corporation, 12277 134th Court NE, Redmond, Washington 98052, USA}
\maketitle

\date{\today}

\section{\label{sec:Intro} Introduction}

A waveguide-fed metasurface consists of one- or two- dimensional arrays of complementary metamaterial elements embedded in one of the conducting surfaces of a waveguide \cite{holloway2012overview,hunt2013metamaterial,Lipworth:13}. Each metamaterial element is excited by the guided wave and radiates a portion of the incident wave into free space. The total radiated field pattern is thus the superposition of the contributions from each of the elements. By adjusting the response of each of the elements---by modifying their geometry or other properties---the field across the aperture can be altered. While there are constraints on the available phase and magnitude associated with tuning the metamaterial elements \cite{smith2017analysis}, the overall performance of the composite aperture\cite{kundtz_microwave_j_2014,smith2017analysis,johnson_ieee_ant_prop_2015,stevenson2016mtenna} can rival that of more complex and costly systems, such as phased arrays and other electronically scanned antennas \cite{williams1981electronically,hansen1966microwave,navarro2008compact,agrawal1999beamformer}, as exemplified by their tremendous success in applications such as terrestrial and satellite communications \cite{stevenson2016mtenna,guerci_microwave_j_2014}, millimeter wave imaging \cite{gollub_nat_sci_rep_2017,yurduseven-gollub-frequency-diverse-human,yurduseven-computational-imaging,Marks:17,sleasman2015dynamic,sleasman_josab_2016,sleasman2017single,diebold2018phaseless} and sensing\cite{del2018dynamic}, and synthetic aperture radar\cite{boyarsky_josaa_2017,pulido_josab_2016,watts2017x,sleasman2017experimental,diebold2017generalized}. 

\begin{figure}
\includegraphics[width=0.9\columnwidth]{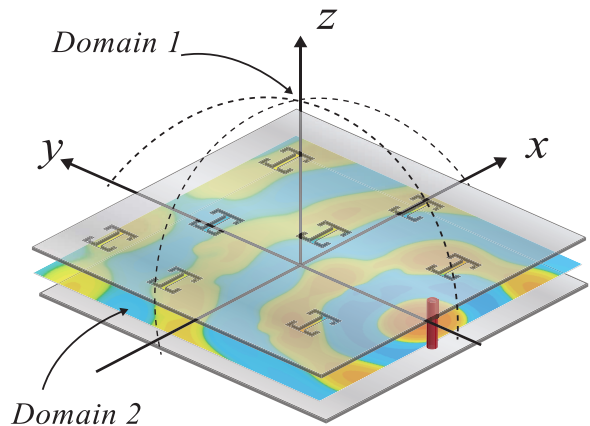}
\caption{Schematic of a waveguide-fed metasurface, divided into two domains of interest.}\label{fig:wg-domains}
\end{figure}

The underlying principles of a waveguide-fed metasurface operation can be understood by conceptually dividing the structure into two domains, as shown in Fig.\ref{fig:wg-domains}. The first domain consists of the half-space outside of the waveguide and shares the radiating surface of the waveguide with the second domain. The field within this half-space can be related to the surface electric and magnetic current distributions over the surface of the waveguide \cite{balanis2016antenna,peterson1998computational}. These current distributions can be virtual, relating to the surface fields using well-known equivalence principles \cite{balanis2016antenna,peterson1998computational}.

The second domain consists of the region within the waveguide, which shares the metamaterial surface with the first domain. A guided mode propagating within the waveguide excites the metamaterial elements, giving rise to a distribution of effective (virtual) current sources over the aperture that then form the radiation patterns in the first domain. Each metamaterial element also scatters back into the waveguide. The field formed on the waveguide-fed metasurface is thus the result of the interaction of the guided wave with each element, as well as the interactions of elements with each other. Accurately capturing these interactions can be vital in predicting the radiated fields, and, consequently, to design desired radiation patterns\cite{eismann1989iterative,sievenpiper2002tunable,sievenpiper2003two,sievenpiper2005forward,fong2010scalar,maci2011metasurfing,pandi2015design,minatti_ieee_ant_prop_2016,johnson_ieee_ant_prop_2015,smith2017analysis}. 


Most analysis techniques proposed for waveguide-fed metasurfaces have relatively limited application and are only useful in special conditions. Commercial full-wave finite element and finite difference solvers are generally intractable given the large contrast between the overall size of the metasurface (many wavelengths) and the metamaterial elements' feature size (usually sub-wavelength). Conventional modeling approaches often rely on the application of quasi-analytical hybrid schemes. Of these, one of the most prominent approaches is that of modulated surface impedances \cite{sievenpiper2002tunable,sievenpiper2003two,sievenpiper2005forward,maci2011metasurfing,minatti_ieee_ant_prop_2016,fong_ieee_ant_prop_2010,patel2011printed,patel2013modeling}. In this framework, the waveguide-fed metasurface is assumed to be composed of a periodic array of metamaterial elements patterned on a grounded dielectric substrate \cite{minatti_ieee_ant_prop_2011,minatti2012circularly}, the collective behavior of which is then modeled as a smoothly varying effective surface impedance with periodic modulation \cite{oliner_ire_trans_ant_prop_1959,oliner1993leaky,sievenpiper2002tunable,patel2011printed,sun2012gradient,pandi2015design}. This approach to model and design metasurfaces was inspired by earlier works on the homogenization of metamaterials \cite{smith2006homogenization,simovski2007local,liu2007description,scher2009extracting}, which was later adapted for metasurfaces\cite{kuester2003averaged,tretyakov2003impedance,tretyakov2003analytical,holloway2011characterizing,tretyakov2015metasurfaces}. Combining this \textit{averaging} model for waveguide-fed metasurfaces with the pioneering works of Oliner and Jackson \cite{oliner1993leaky,jackson2008leaky,oliner_ire_trans_ant_prop_1959}---who considered modulated impedance surfaces as a means of forming desired leaky wave modes---gave rise to a powerful tool for the analysis and synthesis of waveguide metasurfaces. This tool has been successfully used to design a large variety of waveguide-fed metasurface apertures\cite{sievenpiper2002tunable,sun2012gradient,minatti_ieee_ant_prop_2011,minatti_ieee_ant_prop_2015,minatti_ieee_ant_prop_2016,patel2011printed,patel2013modeling,gomez2011analysis,minatti2015modulated,gomez2013holographic,martinez2013holographic,ettorre2012generation,epstein2016huygens,epstein2016cavity}. However, this methodology is primarily applicable when the effective surface impedance is a smooth and gradually varying function of position along the aperture. These averaging techniques cannot be easily extended to arbitrary variations in surface properties. In such scenarios, the surface impedance retrieval process can become complicated as the element geometry exhibits rapid variation and/or loses its symmetries \cite{minatti_ieee_ant_prop_2011,minatti_ieee_ant_prop_2015,maci2011metasurfing,chen2015improving,patel2011printed,patel2013modeling}.

To access the full potential of metasurfaces, it is necessary to alleviate restrictions on element responses and allow arbitrary functionality at each element within the aperture  \cite{sleasman2015dynamic,stevenson2016mtenna,zhu2014dynamic}. Such arbitrary and large variation of elements' responses violates the typical assumptions in homogenization-based models, rendering them ineffective for many aperture profiles of interest. We are thus motivated to formulate a model for the waveguide-fed metasurface that predicts its overall response with minimal restrictions on the constituting elements. Such a model can easily be applied to the previously described modulated surface impedance structures, but, more importantly, can readily be extended to model and design metasurfaces consisting of aperiodic and arbitrary arrangements of metamaterial elements \cite{sleasman2015dynamic,sleasman2016design,haghtalab2018freeform,hsueh2017electromagnetic,memoli2017metamaterial,xie2016acoustic,yang2016programmable}. Foregoing the constrained framework of periodic structures for a more general configuration is especially attractive in the realm of inverse design problems, which have attracted much attention recently \cite{sell2017large,verweij2017objective,wang2018adjoint,marks2018inverse,marks2018linear}. 

To establish a more generalized \textit{analysis} methodology for waveguide-fed metasurfaces, we model each metamaterial element in this paper as a polarizable dipole. This treatment is inspired by a computational technique often referred to as the discrete dipole approximation (DDA) \cite{purcell1973DDA,draine1994DDA,bowen2012using,landy_phot_nano_2013,landy2014two,salary2016model}, in which scattering from a continuous object of arbitrary shape and material properties is modeled as the scattering from a collection of discrete, polarizable dipole scatterers. In the adaptation we present here, we replace each metamaterial element of a waveguide-fed metasurface by an effective magnetic (or electric) dipole moment proportional to the local magnetic field (or electric field), evaluated at the element's location, multiplied by a coupling coefficient termed the \textit{dynamic polarizability}, $\alpha_m(\omega)$ (or $\alpha_e(\omega)$)\cite{scher_meta_2009,karamanos_adv_electromagnetics_2012,landy_phot_nano_2013,pulido2017polarizability}. A simple, first assessment of the radiative properties of such a metasurface can then be found by assuming the guided wave is not perturbed by the metamaterial elements, and each of the elements radiates in proportion to its effective polarizability and does not interact with any other structure \cite{smith2017analysis,lipworth_app_opt_2015,f2016analytical}. Such a simplistic model, while providing considerable intuition, does not lead to accurate field predictions, and thus cannot be used as the basis for a comprehensive design tool. In this paper, we derive analytical formulations to capture the interactions of the elements with each other as well as with the guided wave. Using these formulations, we demonstrate accurate predictions of the field generated by a waveguide-fed metasurface with arbitrary arrangement of metamaterial elements both inside and outside of the waveguide (domains 1 and 2 in Fig. \ref{fig:wg-domains}). 

This paper is organized as follows. In Section \ref{sec:DM}, we review the fundamentals of the coupled dipole modeling. In Section \ref{sec:Derivation} we use the methodology of Section \ref{sec:DM} to extract all the components of the effective polarizability tensors corresponding to two different metamaterial element designs. We also demonstrate that the polarizability is a non-local quantity; i.e. it can be used to find the effective dipole moment in any arbitrary location inside the waveguide. In Section \ref{sec:Fields} we develop an analytical formulation to compute the interaction of metamaterial elements both through the waveguide and through free space. We then compare the predicted scattered fields inside the waveguide using the dipole model with full wave simulations. We also illustrate the importance of capturing the mutual interactions of metamaterial elements to develop an accurate analysis technique. In Section \ref{sec:discussion} we analyze the far field response of a metasurface aperture and demonstrate the role that mutual interactions among the elements play in accurately predicting the radiation pattern. We conclude this paper by examining potential sources of error and propose techniques to improve the accuracy of the methodology. We also discuss the outlook of waveguide-fed metasurfaces and the potential role an accurate analysis methodology can play to accelerate their use in real-world applications. 
\section{Coupled Dipole Model}\label{sec:DM}

\begin{figure}
\includegraphics[width=0.8\columnwidth]{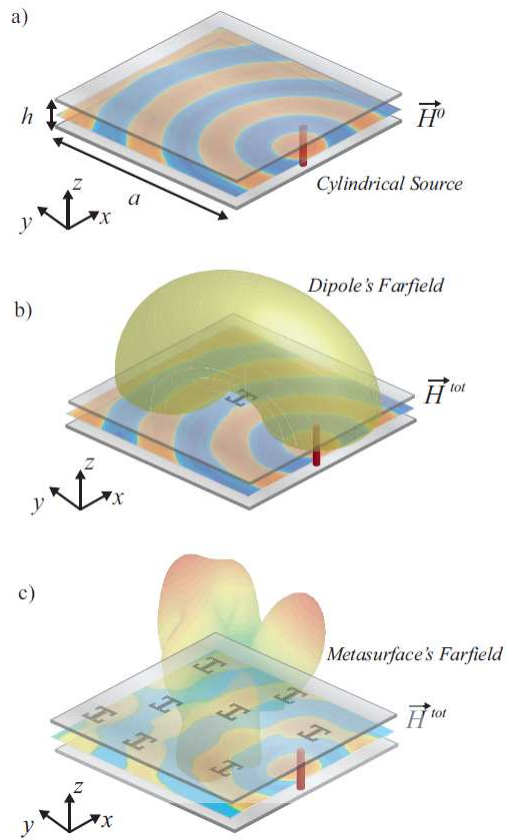}
\caption{Visual steps of coupled dipole model. a) The guided feed generated by a cylindrical source inside the waveguide. Throughout this paper, the dimensions of the planar waveguide are $a=100$ mm and $h=5.21$ mm. b) The interaction of the guided wave and a single metamaterial element, modeled as a magentic dipoles. c) A waveguide-fed metasurface consisting of randomly distributed metamaterial elements. The corresponding overall radiation pattern and the field formed within the waveguide is also depicted.}\label{fig:DDA-explained}
\end{figure}

The steps of the proposed modeling procedure are illustrated visually in Fig. \ref{fig:DDA-explained}. We start from a parallel plate waveguide excited by a probe. This source, assuming the separation of the plates is smaller than half wavelength, launches primarily a cylindrical field inside the waveguide. (Note that a typical coaxial probe possesses more complicated current and near-field distributions, but these can be reasonably ignored in the present analysis and replaced by a simple line source.) Next, we consider the interaction of the cylindrical feed wave with a single metamaterial element. Conceptually, the electromagnetic response of an iris inside a metallic surface is modeled using fictitious magnetic surface currents over the iris, determined by surface equivalence principles \cite{balanis2016antenna,peterson1998computational}. In our proposed model, we assume that each metamaterial element is small compared with the wavelength; therefore, each complementary metamaterial element can be approximated as a magnetic dipole representing the dominant term of the magnetic current density formed over the metamaterial surface \cite{pulido2017polarizability}. This magnetic dipole, $\mathbf{m}$, can be related to the incident magnetic field, $\mathbf{H}_{loc}$ using an effective polarizability tensor, $\bar{\bar{\alpha}}_m$. This relationship is mathematically written as 
\begin{equation}
\mathbf{m}(\mathbf{r}_i)= \bar{\bar{\alpha}}_m \mathbf{H}_{loc}.
\label{eq:mag_dipole}
\end{equation}
\noindent In the case of Fig. \ref{eq:mag_dipole} (b), the field incident on the element, $\mathbf{H}_{loc}$ is equal to the cylindrical feed wave $\mathbf{H}_{0}$. When we have an array of metamaterial elements, as shown in Fig. \ref{eq:mag_dipole} (c), we need to account for the interactions among the metamaterial elements. In this case, the local field acting on the $i-$th dipole (at position $\mathbf{r}_i$) is the superposition of the guided wave at $\mathbf{r}_i$, $\mathbf{H}_{0}$, and the fields at $\mathbf{r}_i$ radiated from all other dipoles on the surface, $\mathbf{H}_{sc}$, or 

\begin{equation}\label{eq:Hloc}
\begin{aligned}
\mathbf{H}_{loc}(\mathbf{r}_i) = \mathbf{H}_{0}(\mathbf{r}_i) + \mathbf{H}_{sc}(\mathbf{r}_i) \\
\mathbf{H}_{loc}(\mathbf{r}_i) = \mathbf{H}_{0}(\mathbf{r}_i) + \sum_{j\neq i} \bar{\bar{G}}(\mathbf{r}_i - \mathbf{r}_j)\mathbf{m}(\mathbf{r}_j) ,
\end{aligned}
\end{equation}

\noindent  where $\bar{\bar{G}}(\mathbf{r}_i - \mathbf{r}_j)$ represents a Green's function. Were these dipoles in free space, the free-space propagator would be the appropriate choice for the Green's function; however, given we are interested in the coupling of the elements through the waveguide, the Green's function is more complicated and we defer specifying the exact form for $\bar{\bar{G}}(\mathbf{r}_i - \mathbf{r}_j)$ until later. A similar equation to Eq. (\ref{eq:Hloc}) can also be written for electric dipoles and the corresponding electric fields. However, for brevity, we have assumed the electric dipole is negligible, as is the case in many situations \cite{pulido2017polarizability}. Examining the expressions in Eq. (\ref{eq:mag_dipole}) and Eq. (\ref{eq:Hloc}), it can be seen that these coupled equations capture the interaction of the incident wave with each of the metamaterial elements (through the $j=0$ terms) as well as interaction between different elements (the summation term). Moving the latter term to the left side of the equation, we obtain a matrix system,

\begin{equation}\label{eq:DDA-eq}
\left\lbrace \bar{\bar{\alpha}}^{-1}_m \delta_{ij} - \bar{\bar{G}}_{ij} \right\rbrace \mathbf{m}_j = \mathbf{H}_{0,i},
\end{equation}
that can be solved for the magnetic dipole moment representing each element, taking into account their mutual interactions. $\delta_{ij}$ is the Kronecker delta function. 

The essential step in transitioning the coupled dipole formalism to a highly accurate modeling tool is the direct computation of the effective polarizabilities of the metamaterial elements. This polarizability extraction step can be performed using full-wave solvers over the domain of a single element, as shown in Fig. \ref{fig:DDA-explained} (b). Since the simulation domain is so much smaller than the entire domain of the composite aperture, this step does not present any significant computational burden, and need only be performed once for a given geometry. The polarizability extraction, combined with the coupled dipole framework, yields an inherently multiscale modeling capability that can be generalized to many types of large-domain antenna architectures.

While previous studies have presented methods to extract the polarizability of a metamaterial element when it is embedded in a waveguide structure \cite{pulido2017polarizability}, no detailed analysis has been presented on the mutual interactions among such metamaterial elements. In this paper, our interest is to clarify the importance of of such interactions, and outline analytical methods to account for them. In the following sections, we will walk though each step of the proposed dipole modeling procedure, highlighting those factors important to accurate field prediction.



\section{\label{sec:Derivation} Metamaterial Element Characterization in Waveguides}
\begin{figure}
\centering{
\includegraphics[width=0.85\columnwidth]{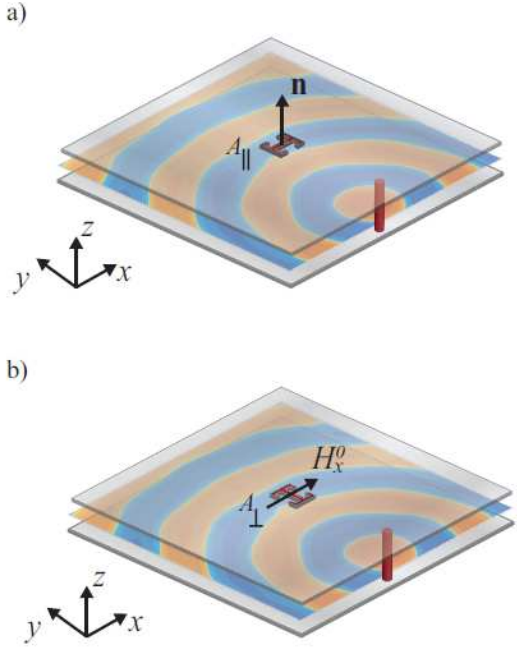}
\caption{Extraction of the polarizability tensor by means of the surface equivalence principle. Given the location of the source and the location of the element, all components of the polarizability tensor can be found following Eqs. (\ref{eq:all-pol-tensor}). a) Setup to extract $\alpha_{xx}$ and $\alpha_{xy}$. c) setup to extract $\alpha_{yx}$ and $\alpha_{yy}$.}\label{fig:pol-tensor}} 
\end{figure}

For the sake of demonstration, we develop the dipole model in a parallel plate waveguide, but the presented methodology can be extended to many other guided-wave structures. To begin, we need to characterize the response of each constituent complementary metamaterial element, which equates to extracting the effective polarizability of each element. Here, we conduct this retrieval process using the surface equivalence principle \cite{pulido2017polarizability,jackson_classical_1999,peterson1998computational,pulido2017extracting}, which states that the fields outside an imaginary closed surface can be fully determined by replacing the closed surface with suitable electric and magnetic current densities that satisfy boundary conditions. Applying this principle to the geometry of the waveguide-fed metamaterial element of Fig. \ref{fig:pol-tensor}, we can determine the field inside and outside the waveguide by replacing the metamaterial element iris with the the effective magnetic surface current given by \(\textbf{M}=\textbf{E}\times\hat{\textbf{n}}\), where $\mathbf{E}$ is the total electric field on the surface of the waveguide and $\hat{\mathbf{n}}$ is the normal to this surface. We note that the tangential electric field is zero everywhere on the waveguide surface except over the void regions defining the metamaterial element (surface highlighted in red). If the element is deeply subwavelength, then the field scattered by the element may be approximated by the first term of the multipole expansions of \(\textbf{M}\). In the first order approximation, the magnetic dipole moment representing the metamaterial element can be calculated as
\begin{equation}
\mathbf{m} = \frac{1}{i \mu_\circ \omega} \int_A \hat{\textbf{n}} \times \mathbf{E}^{tan} da \label{eq:peff-meff}
\end{equation}
\noindent The integration is performed over the surface of the element $A$, $\hat{\textbf{n}}=\hat{\textbf{z}}$ is the vector normal to the top surface, and $\mathbf{E}^{tan}$ corresponds to the tangential field at the surface of the iris. Notice that the magnetic dipole moment $\mathbf{m}$ can be decomposed into $m_x$ and $m_y$. These values are related to the incident magnetic field through a polarizability tensor. At a position where the incident magnetic field has only one component, i.e. $\mathbf{H}_0= H_{0x} \hat{x}$, we can simplify our calculations. As shown in Fig. \ref{fig:pol-tensor} , there are two different scenarios required to find the four components of the polarizability tensor. In the first scenario, cf. Fig. \ref{fig:pol-tensor} (a), the element's main axis is oriented \textit{parallel, $A_\parallel$} with respect to $H_{0x}$ in our coordinate system. We define the \textit{main axis} of a metamaterial element as the symmetry axis providing the largest magnetic response to an applied magnetic field. When the element is excited by the incident field, it induces a magnetic dipole moment $\vec{m}=\{ m_x, m_y\}$, dividing over $H_{0x}$, we obtain the two first components of the polarizability tensor as $\alpha_{xx}= m_x/H_{0x}$ and $\alpha_{xy}= m_y/H_{0x}$. The other components of the polarizability tensor can be found from secondary axis, as shown in Fig. \ref{fig:pol-tensor} (b), where the element is rotated $90^\circ$, i.e. $A_\perp$. To further simplify our calculations, we also utilize the symmetry of the element. If the element possesses mirror symmetry, for example, the extracted polarizabilities satisfy $\alpha_{xy}=\alpha_{yx}$. These considerations can be summarized for the final expressions for the polarizability as (recalling that the incident magnetic field has a single component $H_{0x}$)
 \begin{equation}\label{eq:all-pol-tensor}
 \begin{gathered}
 \alpha_{xx}= \frac{1}{i\mu_\circ \omega H_{0x}} \int_{A_{\perp}} E^{tan}_y da\\
 \alpha_{xy}= \frac{-1}{i\mu_\circ \omega H_{0x}} \int_{A_{\perp}} E^{tan}_x da\\
 \alpha_{yx}= \frac{-1}{i\mu_\circ \omega H_{0x}} \int_{A_{\parallel}} E^{tan}_x da\\
 \alpha_{yy}= \frac{1}{i\mu_\circ \omega H_{0x}} \int_{A_{\parallel}} E^{tan}_y da\\
 \end{gathered}
 \end{equation}
\noindent where the subindex $A_{\parallel}$ corresponds to the element's orientation shown in Fig. \ref{fig:pol-tensor} (a), and $A_{\perp}$ corresponds to the element's orientation shown in Fig. \ref{fig:pol-tensor} (b). 

Following Eq. (\ref{eq:all-pol-tensor}) we extract all four components of the polarizability tensor for two different elements: an elliptical iris, commonly used in slotted waveguide antennas due to its simplicity and single polarization \cite{jackson2008leaky,oliner1993leaky,stevenson1948theory}, and a metamaterial element commonly referred to as a complementary electric inductive-capacitive (cELC) resonator \cite{falcone_prl_2004,schurig2006electric,landy2008perfect,odabasi2013electrically,yoo2016efficient,hand_apl_2008,landy_phot_nano_2013}. The cELC, being a strongly resonant element, is expected to exhibit greater frequency dispersion and thus a narrower region where radiation occurs.

\begin{figure}
\includegraphics[width=0.8\columnwidth]{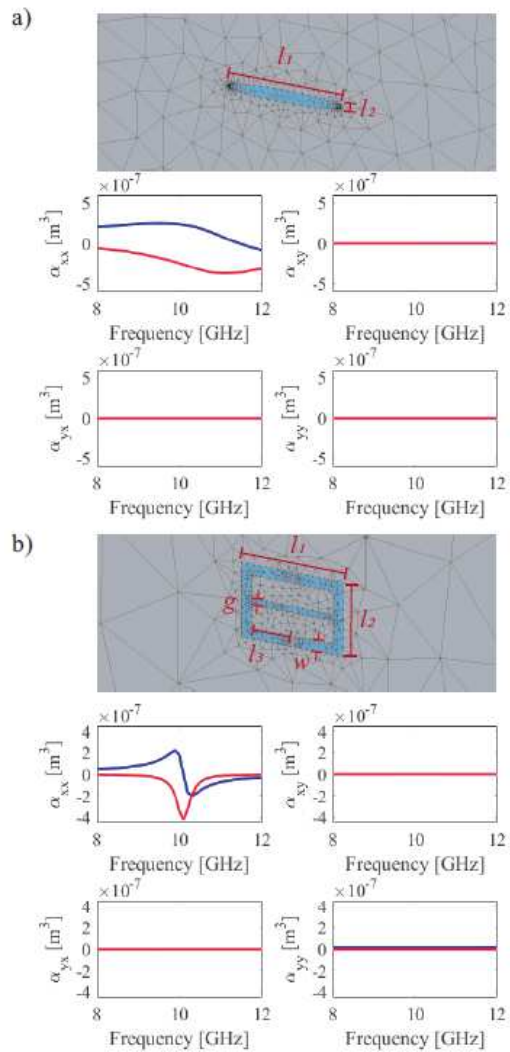}
\caption{Extraction of the polarizability tensor for two different complementary metamaterial elements: a) Elliptical iris. Its dimensions are $l_1= 14.17$mm, $l_2= 1.52$mm b) ELC-resonator. Its dimensions are $l_1=5.26$mm, $l_2=3.5$mm, $l_3=2$mm,$g=0.3$mm, $w=0.51$mm. Accurate meshing of the metamaterial element leads to a more accurate polarizability extraction.}\label{fig:pol-tensor-results}
\end{figure}

We performed a full wave simulation of an air-filled parallel plate waveguide with the elements etched on the top plate, using \textit{CST Microwave Studio}. 
The values of the polarizability tensor across the X-band are calculated for the two elements, as shown in Fig. \ref{fig:pol-tensor-results}. As expected, the elliptical iris shown in Fig. \ref{fig:pol-tensor-results} (a) possesses only one significant component of the magnetic polarizability, $\alpha_{xx}$, which varies slowly over the bandwidth. The ELC shown in Fig. \ref{fig:pol-tensor-results} (b), as expected, exhibits a narrower linewidth. The cELC geometry in this example is selected such that one resonance ($\alpha_{xx}$) appears in the X band; a second resonance associated with the other polarization ($\alpha_{yy}$) appears outside the band of operation. 

\subsection{Independence of the polarizibility on location}
Since the parallel-plate waveguide considered here is translationally invariant, the electromagnetic environment of a given metamaterial element in the absence of any other elements also cannot vary as a function of location. The polarizability is thus an intrinsic parameter of the element and need to be recomputed when the element is shifted to other locations. To illustrate this property, we repeat the calculations above, but with the element placed at a different location. Using the intrinsic polarizability computed for an element at the origin, we multiply by the field at the new location to arrive at the magnetic moments, or
\begin{equation} \label{eq:magnetic-dipole-x0}
\begin{gathered}
m_x= \alpha_{xx}H_{0x}(\boldsymbol{\rho}) + \alpha_{xy}H_{0y}(\boldsymbol{\rho}) \\
m_y= \alpha_{yx}H_{0x}(\boldsymbol{\rho}) + \alpha_{yy}H_{0y}(\boldsymbol{\rho}). 
\end{gathered}
\end{equation}

\noindent We compare this result with full wave simulations that apply the surface equivalence principle, as given by Eq. (\ref{eq:magnetic-dipole-x0}). 

\begin{figure}
\begin{center}
\includegraphics[width=0.9\columnwidth]{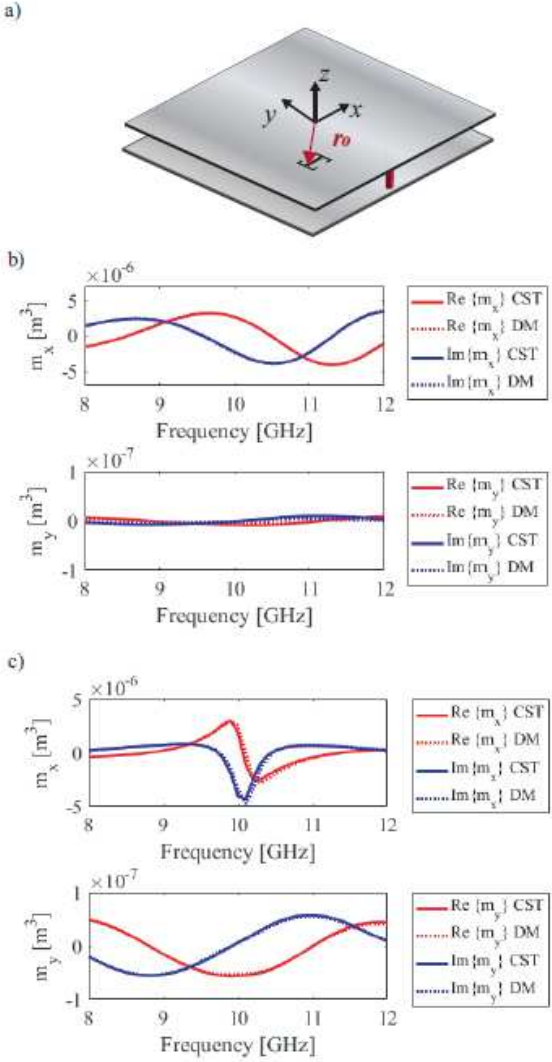}
\end{center}
\caption{Magnetic dipole moment at location $\mathbf{x}_0$. Comparison between the magnetic dipole model as predicted in Eq. (\ref{eq:magnetic-dipole-x0}), and full-wave simulation. a) Diagram of the problem. b) Magnetic dipole moment for the elliptical iris. c) Magnetic dipole moment for the cELC. While $m_y$ is smaller, good agreement between the two methodologies is observed.} \label{fig:mag-dipole-results}
\end{figure}

In our simulation, the excitation is defined as an electric line source of amplitude $I_e=1A$. The incident magnetic field generated by this source is given by the analytical expressions
\begin{equation} \label{eq:H0-theo}
\begin{gathered}
H_{0x}= \frac{iI_e \beta_m}{4}H_1^{(2)}(\beta_m |\boldsymbol{\rho} - \boldsymbol{\rho}'|) \sin \Psi \\
H_{0y}= \frac{-iI_e \beta_m}{4}H_1^{(2)}(\beta_m |\boldsymbol{\rho} - \boldsymbol{\rho}'|) \cos \Psi \\
\end{gathered}
\end{equation}
\noindent where the propagation constant is given by $\beta_m= \sqrt{(2\pi f/c)^2 - (m\pi/h)^2}$, $|\boldsymbol{\rho} - \boldsymbol{\rho}'|$ is the distance from the source to the observation point in the plane of $z=0$, and $\Psi$ is the circumferential angle around the source, measured from $x-$ axis. Figure \ref{fig:mag-dipole-results} presents a comparison between the magnetic dipole moments obtained from the two different methods. As shown, when the element is shifted to the position $\mathbf{r}_0 = (x_0,y_0,0) = (20,20,0)$ $\mathrm{mm}$, excellent agreement between these methods is achieved, demonstrating that the extracted polarizability accounts for the element's intrinsic response at any arbitrary location. In these examples, the element was translated from the origin without rotation; it is important to highlight that if the element undergoes rotation, a rotation matrix should be used over the polarizability tensor before it is replaced in Eq. \ref{eq:magnetic-dipole-x0} to find the total dipole moment. 

\section{\label{sec:Fields} Mutual Interaction Between Metamaterial Elements}
\begin{figure}
\includegraphics[width=0.99\columnwidth]{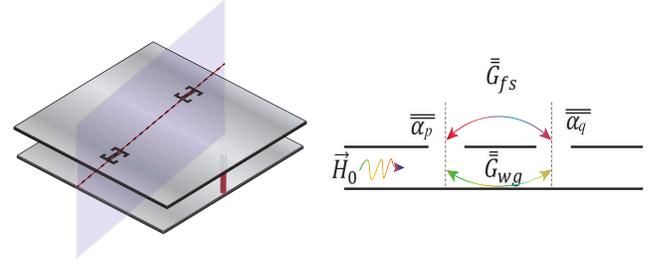}
\caption{Mutual interaction between metamaterial elements. The mutual interaction must account for the propagation of the scattered fields through free-space as well as through free-space.} \label{fig:DDA-mutual-interaction}
\end{figure}
In the previous section we demonstrated that if the element is shifted to an arbitrary location, its total magnetic dipole moment is proportional to its polarizability multiplied by the \textit{incident field} at the element's location. However, in the presence of multiple elements the total field that excites each element is the sum of the incident field plus the scattered field produced by all other dipoles. As shown in Eq. (\ref{eq:DDA-eq}), we can model this interaction in a matrix equation, by appropriately defining the matrices corresponding to the element polarizabilities and the Green's function. With the Green's function technique, a solution of the Helmholtz equation is obtained using an impulse, Dirac-delta as the driving function. For a given problem, the Green's function can take various forms, according to the specified boundary conditions. Considering that the metamaterial elements are patterned at the boundary of two different domains, their mutual interaction, characterized by the Green's function matrix $\bar{\bar{G}}_{ij}$ must account for the analytical expressions of the Green's function in both domains: inside the waveguide $\bar{\bar{G}}^{WG}_{ij}$, and outside the waveguide $\bar{\bar{G}}^{FS}_{ij}$. Inside the waveguide we can describe the components of the Green's function as (See appendix for details):
\begin{subequations}\label{eq:Green-wg}
\begin{align}
G^{WG}_{xx}= \frac{-ik^2}{8h} (H_0^{(2)}(\beta|\boldsymbol{\rho}_p-\boldsymbol{\rho}_q|) \\ 
- \cos (2\boldsymbol{\Psi}_{p,q}) H_2^{(2)}(\beta|\boldsymbol{\rho}_p-\boldsymbol{\rho}_q|)) \nonumber \\
G^{WG}_{xy}= \frac{-ik^2}{8h} \sin (2\boldsymbol{\Psi}_{p,q})  H_2^{(2)}(\beta|\boldsymbol{\rho}_p-\boldsymbol{\rho}_q|)\\
G^{WG}_{yx}= \frac{-ik^2}{8h} \sin (2\boldsymbol{\Psi}_{p,q}) H_2^{(2)}(\beta|\boldsymbol{\rho}_p-\boldsymbol{\rho}_q|)\\
G^{WG}_{yy}= \frac{-ik^2}{8h}( H_0^{(2)}(\beta|\boldsymbol{\rho}_p-\boldsymbol{\rho}_q|) \\
- \cos (2\boldsymbol{\Psi}_{p,q})  H_2^{(2)}(\beta|\boldsymbol{\rho}_p-\boldsymbol{\rho}_q|)).  \nonumber
\end{align}
\end{subequations}
\noindent where $h$ is the height of the parallel plate waveguide, and $|\boldsymbol{\rho}_p - \boldsymbol{\rho}_q|=\sqrt{(x_p-x_q)^2 + (y_p - y_q)^2}$ is the distance between the two dipoles, and $\boldsymbol{\Psi}_{p,q}=\mathrm{tan}^{-1}(\frac{y_p -y_q}{x_p-x_q})$.

Outside the waveguide, the Green's function in free-space is given by
\begin{subequations}\label{eq:Green-fs}
\begin{align}
\bar{\bar{G}}^{FS}(\mathbf{r}, \mathbf{r}')= \left( \frac{3}{k^2 R^2} -\frac{3i}{kR} -1 \right) g(R) \hat{r}\hat{r} \nonumber \\
 + \left( 1+ \frac{i}{kR} - \frac{1}{k^2R^2}\right) g(R) \bar{\bar{I}}  
\end{align}
\end{subequations}
\noindent where $R=|\mathbf{r}-\mathbf{r}'|$, $\bar{\bar{I}}$ corresponds to the identity matrix, $\hat{r}$ and $g(R)= 2\frac{e^{ikR}}{R}$. The factor $2$ in $g(R)$ accounts for the fact that when the elements are seen from the outside of the waveguide, the magnetic dipoles representing the elements are backed by the ground plane and the dipoles have their corresponding self-image. Given the expressions for the dyadic Green's function in Eq. (\ref{eq:Green-wg}) and Eq. (\ref{eq:Green-fs}), it is possible to compute the total magnetic dipole moments by using Eq. (\ref{eq:DDA-eq}) and in turn, compute the scattered fields inside and outside the waveguide. 

\subsection{Waveguide-fed metasurface with arbitrary element arrangement}
\begin{figure*}
\includegraphics[width=0.98\textwidth]{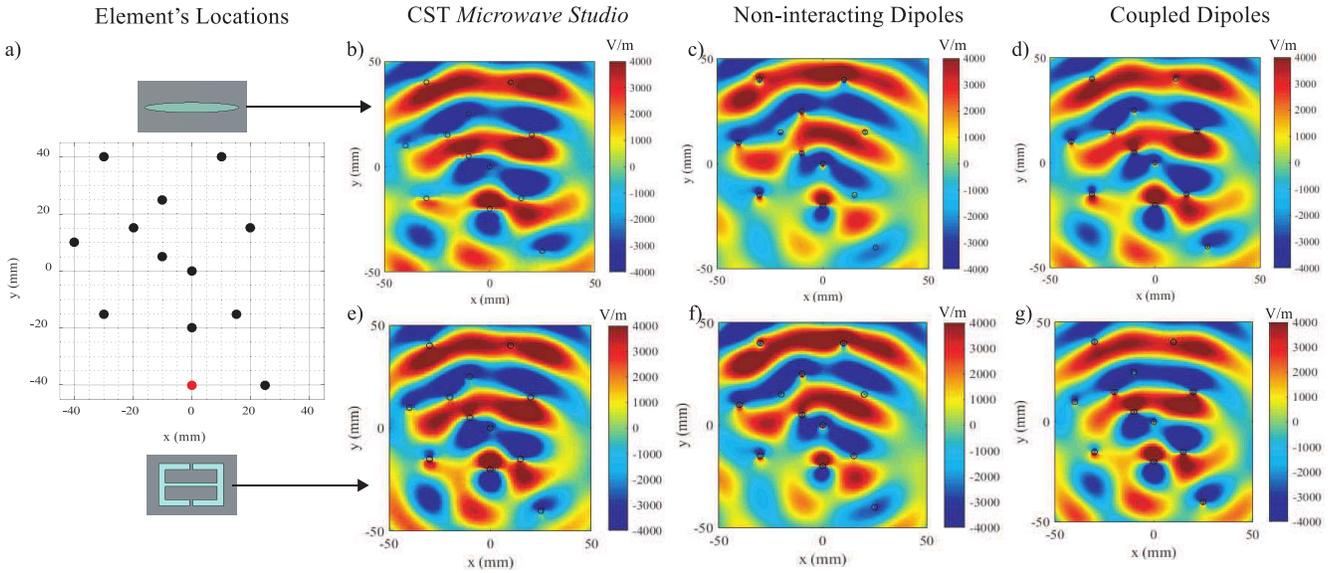}
\caption{Comparison between the scattered fields predicted from full wave simulation and the dipole model. a) Antenna array composed of 12 ELC resonators, red dot represents the source. b) Full wave simulation. c) Dipole model without the mutual element's interaction. d) Dipole model with the mutual element's interaction.}\label{fig:antennaarray}
\end{figure*}

To demonstrate the utility of the model formulated above, we consider a parallel plate waveguide composed of 12 metamaterial elements placed at random locations, as shown in Fig.\ref{fig:antennaarray} (a). It is worth emphasizing such an arbitrary arrangement of strongly resonant metamaterial elements cannot be modeled using any of the conventional methodologies such as modulated surface impedance models \cite{fong2010scalar,minatti_ieee_ant_prop_2011,maci2011metasurfing,fong_ieee_ant_prop_2010,minatti_ieee_ant_prop_2016,sievenpiper2003two,sievenpiper2005forward,pandi2015design,sievenpiper2002tunable,patel2011printed,patel2013modeling}. However, as we will show, by using the dipole model presented here, it is possible to predict with good accuracy the electromagnetic response of such an arrangement, both inside and outside the waveguide. In addition, it is worth noting that this is not a contrived configuration. In fact, such a configuration of arbitrary metamaterial elements have shown great promise in generating frequency diverse patterns used in computational microwave imaging. The model proposed here paves the way for a complete analytical treatment of such structures \cite{hunt2013metamaterial,lipworth_app_opt_2015,Lipworth:13}.

To better illustrate this point, let us examine the interaction effects and impact on the computed field more explicitly. As a means of comparison, we study the scattered electric field inside the waveguide (Domain 1), which will be impacted by the presence the metamaterial elements. The electric field scattered along the $z-$direction generated by a magnetic dipole moment is described by $G^{em}_{zx}$ and $G^{em}_{zy}$. In the presented dipole model (DM), for any arbitrary location of the elements in the waveguide---which might induce both components of the magnetic dipole moments---the scattered $z-$ component of the electric field is then given by
 \begin{equation}\label{eq:E_z}
 E^{DM}_{z,sc}(\boldsymbol{\rho})= \sum _i m_x(\boldsymbol{\rho}_i) G^{em}_{zy}(\boldsymbol{\rho}_i-\boldsymbol{\rho}) + m_y(\boldsymbol{\rho}_i) G^{em}_{zy}(\boldsymbol{\rho}_i-\boldsymbol{\rho}),
 \end{equation}
 \noindent where, as shown in the appendix
 \begin{equation}
 \begin{gathered}
G^{em}_{xz}= \frac{-i\beta^2Z_0}{4h} \left[ H_1^{(2)}(\beta|\boldsymbol{\rho}_i-\boldsymbol{\rho}|)\sin \boldsymbol{\Psi}_i \right] \\
G^{em}_{yz}= \frac{i\beta^2Z_0}{4h}\left[ H_1^{(2)}(\beta|\boldsymbol{\rho}_i-\boldsymbol{\rho}|)\cos \boldsymbol{\Psi}_i \right].
 \end{gathered}
 \end{equation}
\noindent In the above expressions, $Z_0$ corresponds to the free-space impedance. While the expressions for $G^{em}$ remain unchanged, the magnetic dipole moment depends on the mutual interactions.

Figure \ref{fig:antennaarray} (b)-(d) depicts $E_z$ due to the arrangement in Fig. \ref{fig:antennaarray} with the elliptical iris shown in Fig.\ref{fig:pol-tensor-results} (a) as the metamaterial element. Likewise, Fig.\ref{fig:antennaarray} (e)-(g) depicts $E_z$ for the case when the cELC shown in Fig.\ref{fig:pol-tensor-results} (b) is used. Figure \ref{fig:antennaarray} (b) and (e) correspond to the field obtained from a full wave simulation in \textit{CST Microwave Studio}, used here for validation of the model. To highlight the importance of element-element interactions and the utility of the proposed model, we compute $E_z$ using the proposed dipole model in two different ways: first, we ignore the mutual interactions as shown in Fig.\ref{fig:antennaarray} (c) and (f). Next, we include all the mutual interactions, as shown in  and Fig.\ref{fig:antennaarray} (d) and (g). Comparing Figs. \ref{fig:antennaarray} (b), (c), with (d), as well as with Figs.\ref{fig:antennaarray} (e), (f), and (g), significantly closer agreement to full-wave simulations is achieved with the inclusion of the interactions.
In this example, the metamaterial elements have been placed far enough from each other, such that their coupling is only through the fundamental mode of the waveguide. However, in cases where the elements are closer to each other, they may also couple to each other through higher order modes (evanescent modes). We would like to emphasize the framework proposed in this work can also capture such interactions, taking into account that the Green’s function can be expressed as the modal sum of the tensor product of eigenmodes. Therefore, Eq. \ref{eq:Green-wg} can be recast to take into account evanescent guided modes for the mutual interactions. Another important fact to note here is that our model only relies on full-wave simulation of one single element performed only one time. We used the results of that simulation to build a complete model for such a complex metasurface. The presented example of a waveguide-fed metasurface is only $4\lambda$ length, for which full-wave simulation is possible, however, for many applications, much larger aperture with many frequency points need to be examined, and full-wave simulations are not an option. However, our method provides a simple, low cost, and yet effective method to model such complex structures.

\begin{figure}
\centering
\includegraphics[width=0.8\columnwidth]{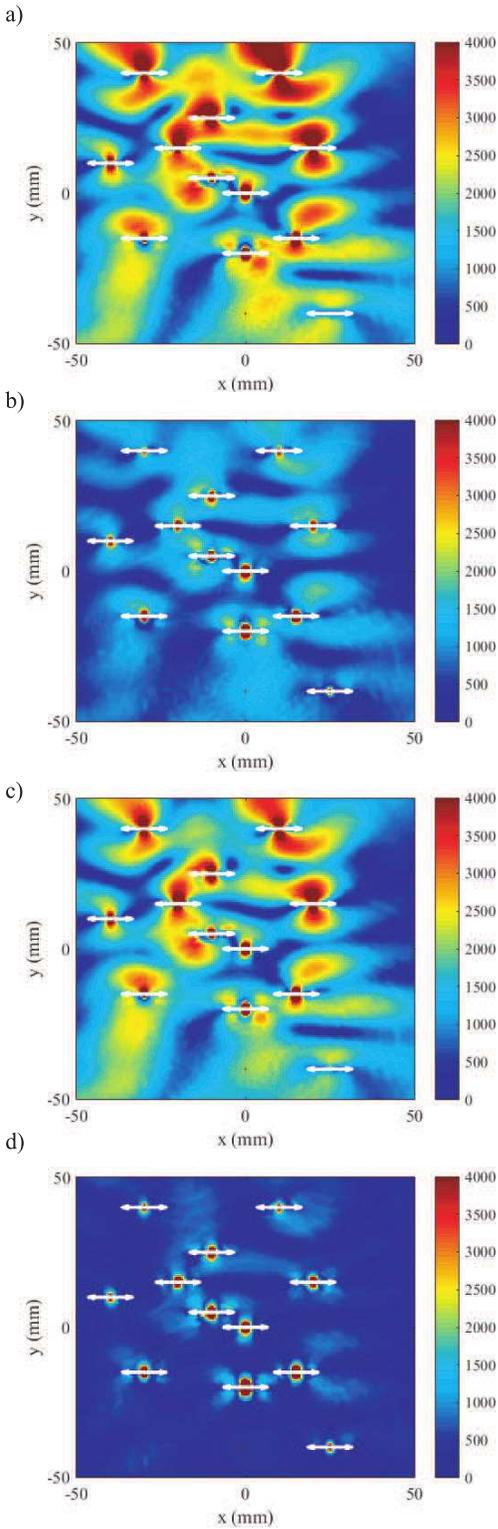}
\caption{Relative error between the scattered field from full wave simulation and the dipole model using 12 elliptical irises. a) Non-interacting dipoles. b) Interacting dipoles inside the waveguide. c) Interacting dipoles in free-space. d) Interacting dipoles across the surface. Arrows represent the elliptical iris' size.}\label{fig:rel-error}
\end{figure}

\begin{figure}
\centering
\includegraphics[width=0.8\columnwidth]{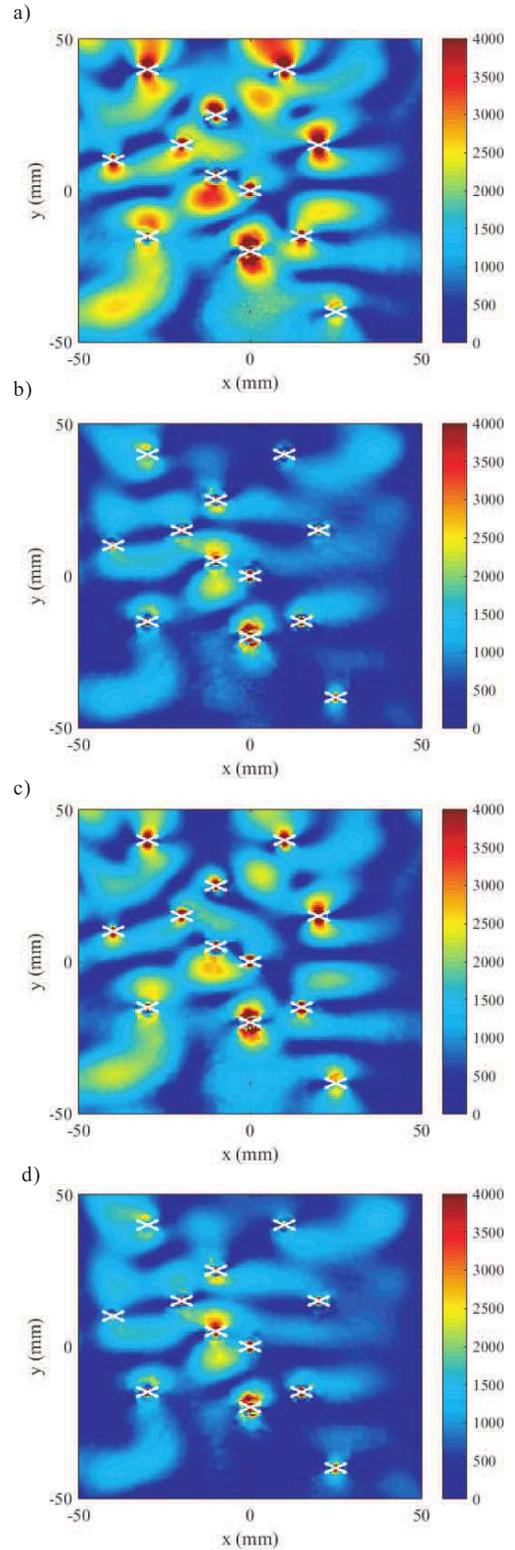}
 \caption{Relative error between the scattered field from full wave simulation and the dipole model using 12 ELC resonators. a) Non-interacting dipoles. b) Interacting dipoles inside the waveguide. c) Interacting dipoles in free-space. d) Interacting dipoles across the surface. Arrows represent the ELC's size }\label{fig:rel-error-elc}
 \end{figure}

While the results presented in Fig.\ref{fig:antennaarray} demonstrate in a qualitative manner the importance of the mutual interaction, we analyze the relative error between the full-wave simulation ($E^{CST}_{z,sc}$) and the dipole model ($E^{DM}_{z,sc}$). More explicitly, this error is defined as $\epsilon= |E^{CST}_{z,sc}-E^{DM}_{z,sc}|$, and is computed in four different scenarios:
\begin{itemize}
\item \textit{Non-interacting dipoles}: In this case, the magnetic dipole moments used in Eq. (\ref{eq:E_z}) result from setting $\bar{\bar{G}}_{ij}=0$ in Eq. (\ref{eq:DDA-eq}). The relative error $\epsilon$ is shown in Fig.\ref{fig:rel-error} (a) and Fig.\ref{fig:rel-error-elc} (a) for the elliptical irises and the ELC resonators respectively. As shown, the relative error is large across the waveguide's domain.

\item \textit{Interacting dipoles inside the waveguide}: The magnetic dipole moments used in Eq. (\ref{eq:E_z}) result from setting $\bar{\bar{G}}_{ij}= \bar{\bar{G}}^{WG}$, as given by Eq. (\ref{eq:Green-wg}), in Eq. (\ref{eq:DDA-eq}). The relative error $\epsilon$ is shown in Fig.\ref{fig:rel-error} (b) for the elliptical irises and Fig.\ref{fig:rel-error-elc} (b) for the ELC resonators respectively. It can be noticed that the error decreases significantly when this mutual interaction is included. Therefore, the element mutual interactions through the waveguide play a significant role in predicting the perturbation of the fields in the waveguide, due to the presence of the dipoles.

\item \textit{Interacting dipoles in free space}: The magnetic dipole moments used in Eq. (\ref{eq:E_z}) result from Eq. (\ref{eq:DDA-eq}) when $\bar{\bar{G}}_{pq}= \bar{\bar{G}}^{FS}$ as detailed in Eq. (\ref{eq:Green-fs}). The relative error $\epsilon$ is shown in Fig.\ref{fig:rel-error} (c) for the elliptical irises and Fig.\ref{fig:rel-error-elc} (c) for the ELC resonators respectively. It can be noticed that the error is reduced compared to the case of no interaction, but can be concluded that it is not the main source of error for the configuration at hand. Therefore, while the mutual element's interaction through free-space is important, it is not as significant as the interaction through the waveguide in the geometries examined here.

\item \textit{Interacting dipoles across the surface}: The magnetic dipole moments used in Eq. (\ref{eq:E_z}) result from Eq. (\ref{eq:DDA-eq}) when $\bar{\bar{G}}_{pq}= (\bar{\bar{G}}^{WG} + \bar{\bar{G}}^{FS} )$. The relative error $\epsilon$ is shown in Fig.\ref{fig:rel-error} (d) for the elliptical irises and Fig.\ref{fig:rel-error-elc} (d) for the ELC resonators respectively. It can be noticed that the relative error is minimum when both interactions are taken into account.

\end{itemize}

Once both mutual interactions are taken into account the lowest error is obtained. Furthermore, since $E^{CST}_{z,sc}$ is the sum of all eigenmodes and $E^{DM}_{z,sc}$ corresponds to the dipolar contribution only, $\epsilon$ can also reveal the role of higher order modes. Figure \ref{fig:rel-error} (a)-(b) shows the relative error $\epsilon$ for the same array of metamaterial elements, using elliptical irises and ELC irises respectively. As shown in Fig.\ref{fig:rel-error} (d), the higher order modes are only noticeable in the close proximity of each element's location and decay rapidly away from the element, demonstrating that the dipolar approximation for the ellipse is quite accurate. However, the dipolar assumption is only a good approximation for the metamaterial elements. While this approximation is sufficient typical scenarios, we intend to fully characterize such interactions in future work, including the impact of higher order multipoles and their mutual interactions.

\section{\label{sec:discussion} Scattered Fields Outside the Metasurface}

\begin{figure}
 \includegraphics[width=0.8\columnwidth]{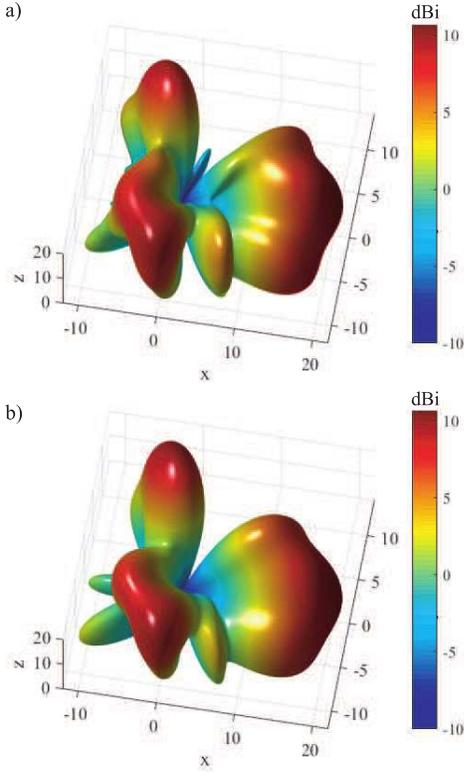}
 \caption{Far-field of the waveguide-fed metasurface composed of cELC resonators. a) Full wave simulation b) Coupled dipole model}\label{fig:farfields3D}
  \end{figure}

\begin{figure}
 \includegraphics[width=0.9\columnwidth]{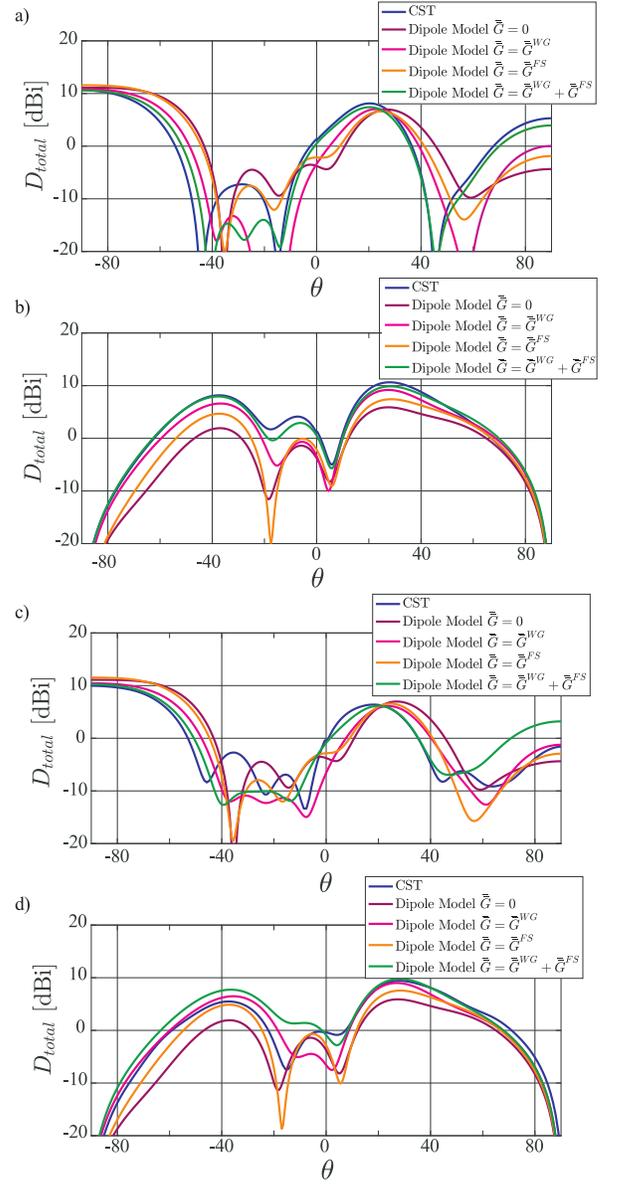}
 \caption{Far-field of the waveguide-fed metasurface in the plane $\phi=0$ :a) for ellipses c) for cELC resonators. And in the plane $\phi=\pi/2$ b) for ellipses d) for cELC resonators}\label{fig:farfield2D}
  \end{figure}

In order to predict the radiated fields from waveguide-fed metasurfaces, we need to solve the problem in domain 1 of Fig. \ref{fig:wg-domains}. In this domain, the metasurface is modeled as an array of magnetic dipoles placed on top of a metallic surface. These magnetic dipoles are computed using Eq. (\ref{eq:DDA-eq}) and include the coupling between elements inside domain 2. In the presented dipole model, it is assumed that the points of observation are in the far-field relative to each dipole. By using the Fraunhofer approximation, $|\vec{r}-\vec{\rho}'|\approx r \; \sqrt[]{1-2\vec{r}\cdot \vec{\rho}'/r^2}$, where $\vec{\rho}'$ is the dipole's location and $r$ the distance of observation, it is possible to separate the radial and the angular dependences of the far-fields. In particular, the radial dependence becomes a simple pre-factor, with the angular distribution of the electric field given by
\begin{equation}\label{eq:E_rad}
 \begin{gathered}
E_\theta \approx \frac{ike^{ikr}}{4\pi r} L_\phi \quad and \quad E_\phi \approx \frac{ike^{ikr}}{4\pi r} L_\theta. 
 \end{gathered}
\end{equation}
The terms $L_\theta$ and $L_\phi$ are defined as
\begin{equation} \label{eq:L_rad}
\begin{gathered}
L_\theta = i\omega \mu \cos \theta \sum^N_{i=1} (m_{xi}\cos \phi + m_{yi}\sin\phi)e^{ik\rho'\cos\psi}\\
L_\phi = i\omega \mu \sum^N_{i=1} (-m_{xi}\sin \phi + m_{yi}\sin \phi)e^{ik\rho'\cos\psi}.
\end{gathered}
\end{equation}

\noindent where $\omega$ is the angular frequency, $m_{x,i}$ and $m_{y,i}$ are the Cartesian components of the magnetic dipole moment for each metamaterial element, and $\theta$ and $\phi$ are the observation angles. In this section, we compute the predicted radiation pattern for the arbitrary arrangement of Fig. \ref{fig:antennaarray} (a). For comparison purposes, we also have computed the radiation pattern using a full-wave simulation.The results are shown in Fig. \ref{fig:farfields3D} where excellent agreement is observed for the radiation pattern at all directions. In addition, in order to analyze the effects of the mutual interaction on the far field pattern, we have compared the directivity at specific planes, i.e. $\phi=0$ and $\phi=90$. As shown in Fig. \ref{fig:farfield2D}, when both mutual interactions are taken into account, the best agreement between full wave simulation and the dipole model (green and blue lines) is obtained. The results in Fig. \ref{fig:farfield2D} once again highlight the capabilities of the proposed method to compute the electromagnetic response of a waveguide-metasurface accurately and efficiently. Here, we did not solve for complex dispersion equations, as is the case in most conventional techniques. Instead, we employ a simple analytical formulation to relate the field of each metamaterial element to the overall response.


 \section{Conclusion and Discussion}
We have presented a robust tool for modeling a waveguide-fed metasurface without minimal restrictions on the metamaterial element or its arrangement. We provide a set of analytical expressions that rigorously capture the interactions among metamaterial elements. The only assumption within the model is that each metamaterial element composing the metasurface is subwavelength and can be reasonably modeled as a point dipole. For translationally invariant geometries, this formulation requires only one full wave simulation of a single element to extract the intrinsic polarizability. Using this intrinsic polarizability, along with analytically derived Greens functions, we can accurately model the response of a waveguide-fed metasurface of arbitrary size or arrangement. Is it important to highlight that the presented calculation of the scattered fields inside and outside the waveguide is computationally inexpensive and allows for the prediction of arbitrary radiation patterns. This capability of the dipole model is particularly interesting for dynamically reconfigurable apertures in which the antenna is composed of highly resonant elements, whose resonance properties can be individual addressed via an external stimulus \cite{sleasman2016design,yoo2016efficient,sleasman2017reconfigurable}.

In particular we demonstrated that an accurate prediction of a waveguide-fed metasurface requires capturing the interaction of metamaterial elements through both the waveguide and free space. Overall, the proposed model, offers a simple, systematic, and computationally low cost method to predict the response of a waveguide-fed metasurface, opening the door to many opportunities: for example, the proposed model can be used to optimize metasurfaces used in computational microwave imaging and replace experimental trial and error \cite{gollub_nat_sci_rep_2017,yurduseven-computational-imaging,yurduseven-gollub-frequency-diverse-human}. It can be used to design and optimize electronically steerable metasurface antennas for communication and imaging applications, where metasurface architectures are now becoming increasingly common.

\section{ACKNOWLEDGMENT}
This work was supported by the Air Force Office of Scientific Research (AFOSR, Grant Nos. FA9550-12-1-0491 and FA9550-18-1-0187). Pulido-Mancera acknowledges support from Kymeta Corporation. 

\appendix
\section{Scattered fields in a parallel plate waveguide}
\begin{figure*}
\begin{center}
\includegraphics[width=0.9\textwidth]{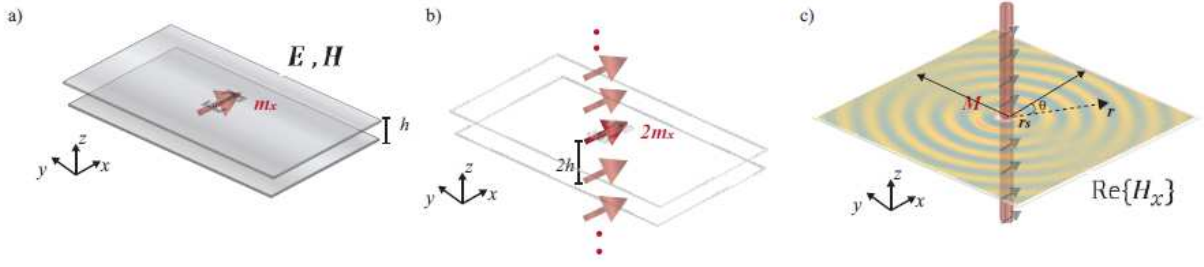}
\end{center}
\caption{Multiple Images of an element embedded in a parallel plate waveguide.}\label{fig:images}
\end{figure*} 
Let us consider a single metamaterial element patterned on the top plate of a parallel plate waveguide. By means of the surface equivalence principle, the element can be represented as a magnetic surface current $\mathbf{M}(\mathbf{r})$. Taking a Taylor expansion over this surface current yields an effective dipole moment given by
\begin{equation} \label{eq:mag_dipole_int}
\mathbf{m}= \frac{1}{j\omega \mu_\circ}\int \mathbf{M}(\mathbf{r}) dV,
\end{equation}
\noindent where $V$ represents the effective volume of the metamaterial element embedded in the waveguide. The effective dipole moment--when observed from the waveguide's domain and applying image theory-- has multiple images along the $z-$axis, as shown in Fig.\ref{fig:images} (a), due to the presence of the plates of the waveguide. Invoking the superposition principles, the magnetic Hertzian potential, $F$, due to this infinite array of magnetic dipoles is given by \cite{landy2013metamaterial}

\begin{equation} \label{eq:hertzian-potential}
\mathbf{F}= \frac{\varepsilon}{4\pi} (j\omega \mu_\circ m_x) \sum^{\infty}_{n=-\infty} \frac{e^{\left(-jk\left[\sqrt{x^2 + y^2 + (z-2nh)^2}\right]\right)}}{\sqrt{x^2+ y^2 + (z- 2nh)^2}}\hat{x}
\end{equation}
where we have assumed the element is along the $x$ direction, and $h$ is the thickness of the substrate.
The convergence of the infinite summation in (\ref{eq:hertzian-potential}) is relatively slow since the terms decay only as a function of inverse radial distance. However, we are interested in the dominant term in the farther distance which can be found by recasting the summation in terms of its discrete spectral components using the Poisson summation technique

\begin{equation}
\sum^{\infty}_{n=-\infty} f(\alpha n) = \frac{1}{\alpha} \sum^{\infty}_{n=-\infty} F \left( \frac{2 n \pi}{\alpha} \right).
\end{equation}

With this technique, the vector potential $\mathbf{F}$ is then,
\begin{equation}
\mathbf{F} =  \frac{\varepsilon}{4\pi h} (j\omega \mu_\circ m_x) \sum^{\infty}_{n=-\infty}  K_0(c_n \rho)e^{jz\frac{2\pi n}{h}}\hat{x}
\end{equation}
\noindent where $c_n= \sqrt{\frac{2 \pi n}{h} - k^2}$, $\rho$ is the axial distance, and $K_0$ is the modified Bessel function of second kind. 
This function is exponentially decaying for real arguments, and propagating for imaginary arguments. Therefore, only the $n=0$ contributes in the far-field. If we only keep the $n=0$ term, then we have the modified Bessel function with $jk\rho$ as its argument. The modified Bessel function can be re-written as
$K_0(jk\rho) = \frac{\pi}{2}(-j)H^{(2)}_{0}(k\rho)$, leading to 
\begin{equation}\label{eq:F-scalarsol}
\mathbf{F}=  \frac{\varepsilon}{4\pi h} (j\omega \mu_\circ m_x)H_0^{(2)}(k\rho) \hat{x},
\end{equation}

The same result can be obtained from a more intuitive approach. When the spacing between the plates are subwavelength, the stack of dipoles can be approximated as an infinite line of magnetic surface current density given by $\mathbf{M^{line}}= M^{line} \delta(\mathbf{\rho})\hat{x}$. The complex amplitude of this magnetic current density, obtained by averaging over the magnetic dipole of (\ref{eq:mag_dipole_int}) over the spacing between them, is given by  
\begin{equation}
M^{line}= \frac{2j\omega\mu_\circ m_x}{2h}.
\label{A2}
\end{equation}
As such, the problem of a single metamaterial element embedded in a parallel plate waveguide can be approximately replaced by the problem of an infinitely long and narrow line of magnetic current density along the $x-$ direction, as shown in Fig.\ref{fig:images} (c). The vector potential for this source can be easily obtained from textbooks, and is given by \cite{harrington1961time}
\begin{equation} \label{eq:vector-F}
\mathbf{F}=  \frac{\varepsilon M^{line}}{4j} H_0^{(2)}(k\rho) \hat{x}
\end{equation}
\noindent which is identical to (\ref{eq:F-scalarsol}) if we substitute expression for $M^{line}$ given in (\ref{A2}) in (\ref{eq:vector-F}).

Once we know the vector potential, $\mathbf{F}$, we can compute the field everywhere. To do that, we first replace
the radial distance in Hankel function argument with $|\boldsymbol{\rho}-\boldsymbol{\rho}_s|$ to represent the distance between the location of the dipole $(x_s,y_s)$ and any point in the plane along the waveguide (x,y), i.e. $\mathbf{\rho}= (x-x_s)\hat{x} + (y-y_s)\hat{y}$. Given the solution for the magnetic vector potential, the scattered magnetic field is
\begin{equation}\label{eq:H}
\mathbf{H}= -j\omega \mathbf{F} - j\frac{1}{\omega \mu_\circ \varepsilon}\boldsymbol{\nabla}(\boldsymbol{\nabla}\cdot \mathbf{F})
\end{equation}
Therefore, the magnetic field is given by 
\begin{equation}
\begin{gathered}
\mathbf{H}= \frac{-\varepsilon M^{line}\omega}{8} \times \\
\left[ H_0^{(2)}(k(|\boldsymbol{\rho}-\boldsymbol{\rho}_s|) - \cos (2\theta) H_2^{(2)}(k(|\boldsymbol{\rho}-\boldsymbol{\rho}_s|)\right] \hat{x}+ \\
 \frac{-M^{line}\omega}{8} \sin (2\theta) H_2^{(2)}(k(|\boldsymbol{\rho}-\boldsymbol{\rho}_s|) \hat{y}.
\end{gathered}
\end{equation}
In order to compute the electric field related to the magnetic vector potential we have
\begin{equation}\label{eq:E}
\mathbf{E}= \boldsymbol{\nabla} \times \mathbf{F}
\end{equation}
Therefore, the electric field is given by
\begin{equation}
\mathbf{E}= \frac{M^{line}k}{4j} H_1^{(2)}(k(|\boldsymbol{\rho}-\boldsymbol{\rho}_s|)\sin \theta \hat{z}.
\end{equation}
 For completion, the expressions for the scattered fields can also be derived for the case when the magnetic dipole is rotated $\pi/2$, i.e. if the surface current is oriented primarily along the $\hat{y}$ direction. In this case, the scattered fields are given by
 \begin{equation}\label{eq:H_dag}
 \begin{gathered}
 \mathbf{H}^{\dagger}= \frac{-M^{line}\omega}{8} \sin (2\theta) H_2^{(2)}(k|\boldsymbol{\rho}-\boldsymbol{\rho}_s|) \hat{x} + \\
 \frac{M^{line}\omega \varepsilon}{8} \left[ H_0^{(2)}(k|\boldsymbol{\rho}-\boldsymbol{\rho}_s|) - \cos (2\theta) H_2^{(2)}(k|\boldsymbol{\rho}-\boldsymbol{\rho}_s|)\right] \hat{y},
 \end{gathered}
 \end{equation}
 \noindent where $.^\dagger$ indicates that the source is oriented along the $\hat{y}$ direction. Analogously, the electric field is
 \begin{equation}\label{eq:E_dag}
 \mathbf{E}^{\dagger} = \frac{M^{line} k}{4j} H_1^{(2)}(k|\boldsymbol{\rho}-\boldsymbol{\rho}_s|) \cos \theta \hat{z}.
 \end{equation}
Once the scattered fields are calculated for two different orientations of the dipole, it is possible to determine all the components of the Green's function as follows:
\begin{equation}
\begin{gathered}
\mathbf{H}= \left\lbrace H_x, H_y \right\rbrace = \left\lbrace m_x G^{mm}_{xx}, m_x G^{mm}_{xy} \right\rbrace \\
\mathbf{H}^\dagger = \left\lbrace H^{\dagger}_x, H^{\dagger}_y \right\rbrace = \left\lbrace m_y G^{mm}_{yx}, m_y G^{mm}_{yy} \right\rbrace, \\
\mathbf{E}={E_z}= m_x G^{em}_{xz}\\
\mathbf{E}^{\dagger}={E^{\dagger}_z}= m_x G^{em}_{xz}
\end{gathered}
\end{equation}
\noindent where $m_x$ and $m_y$ correspond to the magnetic dipole moments that result from the integration of Eq.(\ref{eq:mag_dipole_int}), over the differential volume of the element in the waveguide, and is related to $M^{line}$ via \ref{A2}. Replacing these values of the magnetic dipole moments into Eqs. (\ref{eq:H}), (\ref{eq:E}), (\ref{eq:H_dag}) and (\ref{eq:E_dag}); it is possible to calculate the components of the Dyadic Green's function as
\begin{subequations}\label{eq:Green_mm}
\begin{align}
G^{mm}_{xx}= \frac{-jk^2}{8h} \left[ H_0^{(2)}(k|\boldsymbol{\rho}-\boldsymbol{\rho}_s|) - \cos (2\theta) H_2^{(2)}(k|\boldsymbol{\rho}-\boldsymbol{\rho}_s|)\right] \\
G^{mm}_{xy}= \frac{-jk^2}{8h} \sin (2\theta) H_2^{(2)}(k|\boldsymbol{\rho}-\boldsymbol{\rho}_s|)\\
G^{mm}_{yx}= \frac{-jk^2}{8h} \sin (2\theta) H_2^{(2)}(k|\boldsymbol{\rho}-\boldsymbol{\rho}_s|)\\
G^{mm}_{yy}= \frac{-jk^2}{8h} \left[ H_0^{(2)}(k|\boldsymbol{\rho}-\boldsymbol{\rho}_s|) - \cos (2\theta) H_2^{(2)}(k|\boldsymbol{\rho}-\boldsymbol{\rho}_s|)\right].
\end{align}
\end{subequations}
For clarification, the super-indices in Eq.(\ref{eq:Green_mm}) represent the component of the Green's function that relates the magnetic field due to a magnetic source, and the sub-indices represent the component of the magnetic dipole and magnetic field respectively. For example $G_{mm}^{xy}$ corresponds to the Green's function associated to $H_x$ due to a magnetic dipole oriented along the $y-$direction. Analogously, the electro-magnetic Green's function tensor--relating the electric field due to magnetic sources-- is given by
\begin{subequations}\label{eq:Green_em}
\begin{align}
G^{em}_{xz}= \frac{k^2Z_0}{4h} \left[ H_1^{(2)}(k|\boldsymbol{\rho}-\boldsymbol{\rho}_s|)\sin \theta \right] \\
G^{em}_{yz}= \frac{-k^2Z_0}{4h}\left[ H_1^{(2)}(k|\boldsymbol{\rho}-\boldsymbol{\rho}_s|)\cos \theta \right]
\end{align}
\end{subequations}
\noindent which is proportional to the potential of the magnetic line source shown in Eq.(\ref{eq:vector-F}) \cite{landy2013metamaterial}.
\bibliography{PPWG_DM_references}
\end{document}